\documentclass[11pt]{article}
\usepackage{amsmath,amssymb,color}

\usepackage{amsfonts}

\makeatletter
\@addtoreset{equation}{section}
\makeatother

\newcommand{\hoch}[1]{$\, ^{#1}$}

\newcommand{\be}{\begin{equation}}
\newcommand{\ee}{\end{equation}}
\newcommand{\bea}{\setlength\arraycolsep{2pt} \begin{eqnarray}}
\newcommand{\eea}{\end{eqnarray}}
\newcommand{\nn}{\nonumber}

\def\ben{\begin{equation}}
\def\een{\end{equation}}
\def\half{\frac{1}{2}}
\def\nn{\nonumber}\def\p{\partial}
\def\ft#1#2{{\textstyle{\frac{\scriptstyle #1}{\scriptstyle #2} } }}
\def\fft#1#2{{\frac{#1}{#2}}}

\def\0{{\sst{(0)}}}
\def\1{{\sst{(1)}}}\def\2{{\sst{(2)}}}
 \def\3{{\sst{(3)}}}
\def\4{{\sst{(4)}}}
\def\5{{\sst{(5)}}}
\def\6{{\sst{(6)}}}
\def\7{{\sst{(7)}}}
\def\8{{\sst{(8)}}}
\def\sst#1{{\scriptscriptstyle #1}}

\def\ep{{\epsilon}}

\def\cA{{{\cal A}}}

\begin{document}

\begin{flushright}
\hfill UPR-1258-T
\end{flushright}

\vspace{15pt}
\begin{center}
{\Large {\bf  Quasi-Normal Modes for  Subtracted  Rotating and Magnetised Geometries
}}

\vspace{15pt}
{\Large M. Cveti\v c\hoch{1,3}, G.W. Gibbons\hoch{2,1}
     and Z.H. Saleem\hoch{1,4} }

\vspace{5pt}

\hoch{1}{\it Department of Physics and Astronomy,\\
 University of Pennsylvanian, Philadelphia, PA 19104, USA}

\vspace{5pt}

\hoch{2}{\it DAMTP, Centre for Mathematical Sciences,
 Cambridge University,\\  Wilberforce Road, Cambridge CB3 OWA, UK}
 
 \vspace{5pt}
 
 \hoch{3}{\it Center for Applied Mathematics and Theoretical Physics,\\
University of Maribor, Maribor, Slovenia}

 \vspace{5pt}
 \hoch{4} {\it National Center for Physics, Quaid-e-Azam University,\\
Shahdara Valley Road, Islamabad, Pakistan}

\vspace{10pt}
\underline{ABSTRACT}
\end{center}

{\noindent We obtain explicit  separable solutions 
of the wave equation of massless  minimally coupled scalar fields in the  
subtracted geometry  of  
four-dimensional  rotating and Melvin  (magnetised)  
four-charge black holes of the  STU model, a
 consistent truncation of maximally supersymmetric supergravity with 
four types of electromagnetic fields.  
These backgrounds  possess  a  hidden 
SL(2,R)$\times$SL(2,R)$\times$SO(3)  symmetry and 
faithfully model   the near horizon geometry of  
these black holes, but locate them in a 
confining asymptotically conical box.   
For each subtracted geometry we obtain  
two branches   of quasi-normal modes,  given in terms of hypergeometric functions and spherical harmonics. One branch is  over-damped and 
the other under-damped and   they  exhibit rotational  splitting.  
No black hole bomb is   possible because the Killing field which 
co-rotates with the horizon is everywhere timelike outside the black hole. 
A five-dimensional lift of these geometries is  
given locally  by the product    of a BTZ black hole
with a  two-sphere. This  allows an  explicit analysis of the 
minimally coupled massive five-dimensional scalar field. 
Again, there are   two branches, both damped, 
however now their oscillatory  parts are shifted by the quantised  wave number $k$ along the fifth circle direction. 
\thispagestyle{empty}

\pagebreak

\tableofcontents
\addtocontents{toc}{\protect\setcounter{tocdepth}{2}}

\pagebreak

\section{Introduction}

The  wave equation in the black hole backgrounds provides 
very useful insights into its  internal structure 
and the  relationship with a conformal symmetry 
\cite{Cvetic:1997xv} \cite{Cvetic:1997uw}. The wave equation of a massless scalar field in the background of a general multi-charged rotating black hole turns out to be separable and acquires  an 
SL(2,R)$\times$SL(2,R)$\times$SO(3) 
symmetry, when certain terms are  omitted. In  \cite{Castro:2010fd}   
it was suggested that this symmetry is
 a "hidden conformal symmetry" of the black hole 
that is  spontaneously broken. 

In \cite{CL11I,Cvetic:2011dn}  an explicit example of the  general multi-charged rotating black hole geometry, which exhibits the 
SL(2,R)$\times$SL(2,R)$\times$SO(3) 
conformal symmetry of the  wave equation, was constructed. 
It has been  dubbed  "subtracted geometry" because it is 
constructed by subtracting certain terms from the warp factor of the metric. 
The subtracted geometry preserves the internal structure of the
black hole because it has 
the same horizon area and periodicity  of 
the angular and time coordinates in the near horizon
regions as the original black hole geometry it was constructed from.
The new geometry is asymptotically conical and may  be 
interpreted physically as a black hole in an asymptotically confining box.

This paper is concerned  with subtracted geometries that arise in 
four-dimensional N=2 STU supergravity. This is a consistent truncation of   
maximally supersymmetric ungauged supergravity theories, which arise as an effective theory of toroidally  compactified Type IIA (N=8) or Heteortic (N=4) string theory.  The original four-charge rotating solution \cite{Cvetic:1996kv}\footnote{The full rotating charged black hole seed solution, parameterised by an additional charge parameter was recently obtained in \cite{CC}.},  along with the  explicit expressions for all four gauge potentials was given
in \cite{CCLPII} as a solution of the  bosonic sector  of the
${\cal N}=2$ supergravity coupled to three vector supermultiplets.  
In \cite{Cvetic:2012tr}, it was shown that the corresponding subtracted 
geometry may  be obtained by taking a particular scaling limit of the four-charge rotating black hole solution.  Furthermore, it was shown \cite{Cvetic:2012tr}  that the subtracted geometry for the Schwarzchild black hole can be obtained by applying  Harrison transformations of the STU model, 
which comprise a part of the larger set 
of  symmetries of the black holes when the four-dimensional
black hole Lagrangian is  
reduced on time to  three dimensions. In \cite{Vir,Cvetic:2013cja,Sahay:2013xda}, 
this was generalized to the case of four-charge rotating black holes of the 
STU model and the interpolating solutions  between the 
rotating black holes and their  subtracting geometry   
were obtained  \cite{Cvetic:2013cja,Jottar} by continuously  varying the 
 boost  parameters of the Harrison transformations from zero to an infinite boost.  
(For related works on extremal subtracted geometries, see  \cite{Chakraborty:2012nu,Chakraborty:2012fx}.)

This paper  will also be dealing with the subtracted geometry  of  Melvin STU black holes 
which  arise as a  scaling limit of   magnetised four-charge black holes. 
The  magnetised four charge black holes of the STU model were  constructed in 
\cite{Cvetic:2013roa} by solution generating techniques.  Special cases include   Schwarzschild and Reissner-Nordstr\"om black holes 
in the magnetic field of Maxwell-Einstein gravity.   
The subtracted geometries  of Melvin STU black holes were
 also constructed there. These geometries 
faithfully model the  near-horizon region of  
multi-charged black holes in magnetic field backgrounds. 

The physical properties of the black holes in the magnetic backgrounds can typically be studied only numerically. We shall see that these  magnetised backgrounds can be analysed analytically.

The main aim of this paper is to analyse the quasi-normal solutions of the scalar wave equation in the background of the above mentioned  subtracted rotating  geometry and the subtracted magnetised geometry, by employing their  hidden 
SL(2,R)$\times$SO(2,R)$\times$SO(3)  symmetry. We do so by   
first explicitly solving the wave equation for a massless scalar field in 
four dimensions,  which due to the very special structure of the 
metric is separable and solvable   in terms of 
 hypergeometric functions and spherical harmonics 
both for  subtracted rotating and   subtracted  magnetised geometries.  
In each case we obtain two branches of quasi-normal modes, 
with remarkably simple  values of complex eigenfrequencies,  
one over-damped and one under-damped.  
Specifically,  in the   case of magnetised geometries the 
effect of the magnetic field  turns out to  be an additive 
 shift of  the real part of the eigenfrequency of the quasi-normal modes.  
The regularity of these  solutions near the outer horizon is  
analysed in terms of Kruskal-Szekeres coordinates. 
These results are presented  for subtracted rotating   geometries 
 in Section 2 and  for subtracted magnetised geometries  in Section 3.

The analysis is further extended by studying the wave equation for a 
minimally coupled massive scalar field in the  five-dimensional lift 
of these  subtracted geometries.  For both  rotating and  magnetised cases, 
the lift on a circle  $S^1$ results in  a  geometry that is locally  ${\rm  BTZ }\times S^2$, a product of the BTZ black hole and a two-sphere.
 As a consequence,  the wave equation for a 
massive minimally coupled scalar field is separable and  may 
 be solved again in terms of the hypergeometric functions,  
spherical harmonics and a plane wave along the $S^1$ circle direction.  
Remarkably simple, explicit  expressions for the 
frequencies of the two branches of the quasi-normal modes are obtained , 
where the quantised wave number along the  $S^1$ circle    
shifts the real part of 
he eigenfrequencies. For the special case of  the 
zero wave number and zero five-dimensional mass, 
one reproduces the  results of 
Sections 2 and 3  as expected.  
Solutions for the non-zero wave numbers can be interpreted as 
quasi-normal modes for the  massive  four-dimensional Kaluza-Klein modes  
whose  electric charge is proportional to the wave number. The regularity of these modes near the outer horizon is manifest after  performing  a Kaluza-Klein U(1) gauge transformation  on the wave function.  
All of   these results are presented in Section 4.

The paper also contains a number of Appendices (Section 5)
collecting together results needed for the calculations described above
in a uniform notation. 
 Section  5.1 provides explicit formulae for  
the subtracted rotating geometry and all the fields of the STU model,  
which were worked out in  \cite{Cvetic:2012tr}.   
Section 5.2 does the same for the   subtracted magnetised  geometry in the STU model  by elaborating on  results given in \cite{Cvetic:2013roa}.
Section 5.3 gives detailed expressions for the lift of  these geometries 
on a circle to five dimensions, leading to the 
${\rm BTZ}\times S^2$ geometry. Earlier partial results for the 
rotating geometry were given in \cite{Cvetic:2011dn,Cvetic:2012tr}, 
and for the magnetised  one in \cite{Cvetic:2013roa}. 
Here  particular care  is taken of the dimensions  
and  of  the periodicities of metric coordinates.   
In Section 5.4 a map is provided  taking  the BTZ coordinates to 
the local AdS$_3$ metric from \cite{Banados:1992wn,Banados:1992gq}. 
 Section 5.5  contains the formulae for the   
Kalulza-Klein reduction of the scalar wave equation on a circle.

\section{ Subtracted Rotating Geometry}

The metric for the four-charge rotating black hole solution of the STU model  can be written in the form \cite{Cvetic:1996kv,CCLPII,Cvetic:2011dn}:
\be
d  s^2_4  = -  \Delta^{-\frac{1}{2}}_0  G  
( d{ t}+{ {\cal  A}})^2 + { \Delta}^{\frac{1}{2}}_0 
(\frac{d r^2} { X} + 
d\theta^2 + \frac{ X}{  G} \sin^2\theta d\phi^2 ),\label{metric4d}
\ee
with 
\bea
{ X} & =& { r}^2 - 2{ m}{ r} + { a}^2~,\cr
{ G} & = &{ r}^2 - 2{ m}{ r} + { a}^2 \cos^2\theta ~, \cr
{ {\cal A}} & \equiv & \frac{{a\sin^2\theta A_{red}}}{G}=\frac{2{ m} { a} \sin^2\theta}{ G}
\left[ ({ \Pi_c} - { \Pi_s}){  r} + 2{ m}{ \Pi}_s\right] d\phi~,\label{c4d}
\eea
and the warp factor $\Delta_0$  given by
\bea
{ \Delta}_0 =&& \prod_{i=1}^4 ({ r} + 2{ m}\sinh^2 { \delta}_i)
+ 2 { a}^2 \cos^2\theta [{ r}^2 + { m}{ r}\sum_{i=1}^4\sinh^2{ \delta_i}
+\,  4{ m}^2 ({ \Pi}_c - { \Pi}_s){ \Pi}_s  \cr && -  2{ m}^2 \sum_{i<j<k}
\sinh^2 { \delta}_i\sinh^2 { \delta}_j\sinh^2 { \delta}_k]
+ { a}^4 \cos^4\theta\, .
\eea

The mass, four charges and the angular momentum are parameterised as
\bea
G_4 M  &= &\frac{1}{4}m\sum_{i=1}^4\cosh 2\delta_i\, ,\nn\\
G_4 Q_i & = &\frac{1}{4}m\sinh 2\delta_i\, ,\quad i=1,2,3,4\, ,\nn\\
G_4 J & = &m a (\Pi_c - \Pi_s)\, , 
\eea
with $G_4$ the four-dimensional Netwon's constant and we employ the abbreviations
\be
\Pi_c \equiv \prod_{i=1}^4\cosh\delta_i
~,~~~ \Pi_s \equiv  \prod_{i=1}^4 \sinh\delta_i~.
\ee
The two horizons, given  by $X=0$, are at
\be
r_\pm=m\pm \sqrt{m^2-a^2}\, .
\ee
 It was shown in  \cite{Cvetic:2011dn}  that the replacement 
\be
\Delta_0 \rightarrow  \Delta =  (2m)^3r(\Pi_c^2-\Pi^2_s) +(2m)^4\Pi_s^2
-(2m)^2 (\Pi_c-\Pi_s)^2 a^2 \cos^2 \theta \,,  \label{warpf}
\ee
in  the metric (\ref{metric4d}) reduces  the highest  
power of $r$ in $ \Delta_0$
and renders in the radial part of the  massless scalar wave equation  the irregular singular point at infinity regular, allowing for
solutions in terms of hypergeometric functions.
Moreover, the massless scalar wave equation is  separable
in terms of ordinary spherical harmonics, rather than the complicated
spheroidal functions needed for the  full  four-charge black hole solution.  
This new metric has been  dubbed a ``subtracted geometry''  and 
the massless scalar wave equation    in this background
exhibits a hidden SL(2,R)$\times$SL(2,R)$\times$SO(3)
symmetry.  Furthermore, at the outer and inner horizons the entropies
\be
S_{\pm}  = \frac{2\pi m} {G_4}\left[ (\Pi_c + \Pi_s)m \pm (\Pi_c - \Pi_s)\sqrt{m^2-a^2} \right]\, ,
\ee
the inverse  surface gravities
\be
\frac{1}{\kappa_{\pm}}=2m\left[ \frac {m}{\sqrt{m^2-a^2}}(\Pi_c + \Pi_s) \pm (\Pi_c - \Pi_s)\right]\, ,\label{kappa}
\ee
and  the angular velocities
\be
\Omega_{\pm} =\kappa_{\pm} {a \over\sqrt{m^2 - a^2}}~,\label{angular}
\ee
remain unchanged by this replacement,  thus
preserving the  local geometry and thermodynamic properties of the metric. 
The expressions simplify significantly in the static case when $a=0$.

It is straightforward to see that these black hole solutions and their subtracted geometry encompasses  the following  special cases:
\bea
\hbox{Kerr-Newman:}&& \quad \delta_1=\delta_2=\delta_3=\delta_4\, , \nn\\
\hbox{Kerr:}&&\quad \delta_i=0\, , \quad i=1,2,3,4\, ,\nn\\
\hbox{Reissner-Nordstr\"om:}&&\quad \delta_1=\delta_2=\delta_3=\delta_4\, , \quad  a=0\, , \nn\\
\hbox{Schwarzschild:}&&\quad \delta_i=0\, ,  \quad  a=0\, ,\quad i=1,2,3,4\,  .
\eea
 
\subsection{Kruskal-Szekeres  Coordinates  for  Subtracted Rotating Geometry}
In the following we construct Kruskal-Szekeres type  coordinates to cover
the outer horizon which allow us to identify suitable
boundary conditions there\footnote{One can analogously construct  Kruskal-Szekeres type  coordinates to cover
the  inner horizon  region.}. At infinity the 
appropriate boundary condition is boundedness of the solution. 
The construction of Kruskal-Szekeres  coordinates is in fact considerably simpler than that
used for the Kerr solution \cite{Carter1,Carter2}.

The subtracted metric (\ref{metric4d}), (\ref{c4d}) with (\ref{warpf}) can be
 cast in  the following  remarkably simple form\footnote{This structure was also anticipated in \cite{Cvetic:2011dn} by evaluating the Laplacian of the subtracted rotating geometry.}:
\be
ds ^2 = \sqrt{\Delta} \frac{X}{F^2}
 \Bigl (- dt ^2 + \frac{F^2 dr^2} {X^2 } \Bigr )+
  \sqrt{\Delta} d\theta ^2 + \frac{F^2 \sin ^2 \theta}{ \sqrt{\Delta}  }  (d \phi + W d t) ^2  \, , \label{metricn}
\ee
with
\be
W= -\frac{aA_{red}} {F^2} \, , \quad
F^2 =(2m)^2\left[2m ( \Pi _c^2 - \Pi _s ^2 ) r  +(2m)^2 \Pi_s^2 -a^2 (\Pi_c-\Pi_s)^2\right].
\ee
$X$ and $A_{red}$ are  defined in (\ref{c4d}) and we display them again
\be 
X=r^2-2mr +a^2\, , \quad A_{red}=2m(\Pi_c-\Pi_s)r +(2m)^2 \Pi_s\, .
\ee
Importantly, $X$, $F$ and $W$ are only functions of $r$. We also note that 
 the  factor $\Delta$  (\ref{warpf}) can be written in terms of $F^2$ as
\be
\Delta=F^2 + (2m)^2a^2(\Pi_c-\Pi_s)^2 \sin^2\theta\, .
\ee
It is straightforward to show that
\be\frac{1}{\kappa_{\pm}}=\frac{2 F(r_\pm)}{r_+-r_-}\, ,   \label{Fp}
\ee
and
\be 
\Omega_{\pm}= -W(r_\pm)\, . \label{Omega}
\ee
This special property of  the angular velocities and 
surface gravities leads to an asymmetry of two branches of the 
quasi-normal modes as analysed later in this Section.

We now construct Kruskal-Szekeres type  coordinates to cover
the horizon which allow us to identify suitable
boundary conditions there. 
Due to the structure of the metric (\ref{metricn}) the construction of Kruskal-Szekeres  coordinates is straightforward.

The metric (\ref{metricn})  allows for the introduction of 
retarded and advanced co-rotating  Eddington-Finkelstein  
coordinates:
\be
u= t- r* \,, \qquad v= t+ r* \,,\qquad   \phi_+ = \phi  + W(r_+) t  \, ,\label{EF}
\ee
which satisfy
\be
g^{\alpha \beta } \p _\alpha u \p_  \beta u  = 0 =
 g^{\alpha \beta } \p _\alpha v \p_  \beta v \, .
\ee
The Hamilton-Jacobi equation is separable, yielding a solution
\be
r* =  \int ^r \frac{Fdr}{X}\, ,
\ee
which is  manifest for the metric (\ref{metricn}). 

The co-rotating Kiling vector
\be
l^+=\frac{\partial}{\partial t} -W(r_+) \frac{\partial}{\partial \phi}\, , \label{lp}
\ee
 coincides 
with the null generator of the horizon.  The angle $\phi_+$ is constant along
the orbits of the co-rotating Killing vector $l^+$:
\be
l^+  \phi_+ = (\p_t -W(r_+) \p_\phi) \phi_+ =0\,. 
\ee
We introduce Kruskal-Szekeres coordinates:
\be
U= - e^{-\kappa_+ u} \,, \qquad V=  e^{\kappa_+ v}\, ,\label{KS}
\ee 
and thus
\bea 
 \frac{dV}{V}+\frac{dU}{U}&=&\frac{  2 \kappa _+ F dr}
{X}  \, , \nn\\
\frac{dV}{V}-\frac{dU}{U} &=&
 2 \kappa_+ dt\, .\label{UV}
\eea
In terms of Kruskal-Szekeres coordinates the  metric (\ref{metricn}) takes the following form:
\bea
ds ^2 &=& \sqrt{\Delta}\, \frac{X}{F^2}\frac{\, \, \, dU dV}{\kappa_+^2 U V}\, +
     \sqrt{\Delta} d\theta ^2\, \nn\\
      &+& \frac{F^2\sin^2\theta}{\sqrt\Delta  } \left[d \phi_+ +
 \frac{1}{2\kappa_+} (W(r)-W(r_+) )(\frac{dV}{V}-\frac{dU}{U}) \right] ^2  \, . \label{metricnn}
\eea
In the vicinity of the outer horizon $r\sim r_+$ one has
\be
r*=\int ^r \frac{F(r) dr}{X}  \sim
\frac{F(r_+)}{r_+-r_-} \ln (r-r_+) =\frac{1}{2\kappa_+}\ln(r-r_+) \, , 
\ee
where we used (\ref{Fp}) at the last step.  This ensures
\be
-UV= e^{2\kappa_+r*} \sim (r-r_+) \, ,   
\ee
and the metric  (\ref{metricnn})  is regular and analytic.

An argument given by Hawking and Reall \cite{Hawking:1999dp} 
in the asymptotically AdS  case may be adapted to  show that
if the co-rotating Killing vector $l^+$  (\ref{lp}) is
timelike outside the horizon then there can be no super-radiance
instability or  a black hole bomb \cite{Press:1972zz,Cardoso:2004nk}.

The length squared of the co-rotating Killing vector $l^+$ (\ref{lp}) is
\be
g^{\alpha \beta }l^+ _\alpha   l^+ _\beta= -\frac{1}{\sqrt{\Delta}}\left[X+\frac
{a^2\sin^2\theta  (\Pi_c-\Pi_s)^2(r_+-r_-)(r-r_+)}{\left[(\Pi_c-\Pi_s)r_++2m\Pi_s\right]^2}\right]
\, . \ee
which is manifestly  negative for $r>r_+$ and thus their is no super-radiance.

\subsection{Massless Wave Equation and Quasi-Normal Modes}

The massless scalar wave equation  for the multi-charge black hole metric  (\ref{metric4d}) is separable
and the solutions  expressible  in terms of spheroidal functions
of $\theta$  \cite{Cvetic:1997uw,Cvetic:1997xv}. 
The  radial function may be expressed in terms of solutions of  
a confluent form of Heun's
equation which has two regular singular points   and an 
irregular singular point at infinity. 

For the subtracted geometry metric (\ref{metricn}) 
the massless scalar wave equation is also separable and of a specific form:

\be 
e^{-i\omega t} e^{in \phi} P^n_l(\theta) \chi(x) 
\, , \label{ansatz}
\ee
where  $P^n_l(\theta)$ is an associated Legendre 
polynomial, the solution of the  unit two-sphere $S^2$ Laplacian with eigenvalues $l(l+1)$,   $l=0,1,\dots$ and $n=\pm l, \pm(l-1), \dots$

The radial equation takes the form
\cite{Cvetic:1997uw,Cvetic:1997xv}:  
\be
\Bigl[ \frac{\partial}{\partial x} (x^2 - 
\frac{1}{4})\frac{\partial}{\partial x} +  
\frac{1}{4(x-\half )}\bigl( \frac{\omega}{\kappa_+} - 
n\frac{\Omega_+}{\kappa_+} \bigr)^2 
  - \frac{1}{4( x+\half)}
 \bigl(\frac{\omega}{\kappa_-}  -
 n \frac{\Omega_-}{\kappa_-} \bigr)^2 -  l(l+1) \Bigr] \chi(x)=0\, , \label{wave}
\ee
where 
\be
x = \frac{r - \frac{1}{2}(r_+ + r_-)} { r_+ - r_-}~,\label{dlr}
\ee
is designed so that the  two horizons $r_\pm$ are at   
$x=\pm \frac{1}{ 2}$ . 

Due to  (\ref{angular}) rotating solutions have the property:
\be \frac{\Omega_+}{\kappa_+}=\frac{\Omega_-}{\kappa_-}\, , \label{ratio}
\ee
 and thus the solutions to (\ref{wave})  depend only on one ratio $\Omega_+\kappa_+^{-1}$, only.

Solutions which are ingoing on the future horizon
must be regular at $U=0$ in Kruskal-Szekeres coordinates and
this implies  \cite{Cvetic:1997uw,Cvetic:1997xv,Cvetic:2011dn}    
\bea
\chi(x)&=& (x+\half)^{- (l+1) }
\bigl( \frac{x-\half}{x+\half } \bigr ) ^{ -i (\omega - n \Omega _+) \frac{\beta_H }{4 \pi }}\nn \\
&\times&F ( l+1-  i  \frac{\beta_R \omega -2n \beta _H \Omega _+}{4 \pi},
 l+1-  i  \frac{\beta_L\omega }{4 \pi}, 1-   i \frac{\beta_H (\omega - 
n \Omega _+ ) }{2 \pi }
; \frac{x-\half}{x+\half } ) \,,  \eea
where 
\ben
\frac{\beta _H}{2 \pi} = \frac{1}{\kappa_+} \,, \quad 
 \frac{\beta _R}{2 \pi } = \frac{1}{\kappa_+} + \frac{1}{\kappa _-} \,,\quad 
\frac{\beta _L}{2 \pi } = \frac{1}{\kappa_+} - \frac{1}{\kappa _-}\,.  
\een
Near the outer horizon  $r^\star \rightarrow - \infty,\,
(x-\half)(x+\half )^{-1} \rightarrow e^{2 \kappa_+ r^\star } 
$  and so 
\be
 \chi(x) \approx 
e^{-i  (\omega - 
n \Omega _+)  r^\star} F( l+1-  i  \frac{\beta_R \omega -2n \beta _H \Omega _+  }{4 \pi},
 l+1-  i  \frac{\beta_L\omega }{4 \pi}, 1-   i \frac{ \beta_H(\omega  - n \Omega _+)  }{2 \pi }
; e^{2 \kappa _+ r^\star}  )\, .
\label{ingoing} \ee
In Kruskal-Szekeres coordinates therefore
\ben
e^{-i\omega t} e^{in \phi} \chi (x) \approx e^{in \phi_+ } 
 V^{-i\frac{\omega - n \Omega_+}{\kappa _+ } }  ( 1 + \dots )   \, , 
\een
where the ellipses denote a power series in $UV$
which is convergent in a neighbourhood of the future horizon 
$U=0$ \,.  

At large $x$  \cite{Cvetic:1997uw,Cvetic:1997xv}
\bea
\chi(x)& \approx&  x ^{-(l+1)}
\frac{
\Gamma (1 -i  \frac{\beta_H (\omega -n  \Omega _+) }{2 \pi }) \Gamma (-2l-1 ) }
 {\Gamma (-l -i \frac{\beta_L \omega}{4 \pi})
 \Gamma (-l -i  \frac{\omega \beta_R - 2n \beta _H \Omega _+ }{4 \pi}  )
}\nn\\
&+& x^l  
\frac{\Gamma (1 -i  \frac{\omega \beta_H (\omega -n\Omega _+)  
}{2 \pi }) \Gamma (2l  +1 )} 
{\Gamma ( l+1 - i  \frac{\beta_L\omega}{4 \pi}) 
\Gamma (l+1 -i   \frac{\omega \beta_R - 2n \beta _H \Omega _+ } {4 \pi}) 
} \, . 
\eea
In order that $\chi$ be finite at spatial infinity, we must set  
\bea
&&i \omega  \frac{\beta_L}{4 \pi} = l+1 +N_L
\,, \nn \\
\hbox{or}\quad
&&i   \frac{\omega \beta_R - 2n \beta _H \Omega _+}{4 \pi}
=l+1 +N_R \,, 
\eea
where $N_{L,R}=0,1,\dots$
This gives remarkably simple  formulae for the frequencies of the quasi-normal modes
\bea
&&\omega=-\frac{i}{2m(\Pi_c-\Pi_s)}(1+l+N_L)\, , \nn \\
\hbox{or}\quad &&\omega=-\frac{i\sqrt{m^2-a^2}}{2m^2(\Pi_c+\Pi_s)}(1+l+N_R)+ \frac{a}{2m^2(\Pi_c+\Pi_s)}n\,. \label{qmode}
\eea
  Both  frequencies result in damped modes, with the under-damped branch exhibiting oscillatory behaviour and the  damping absent  in the extremal limit $a\to m$. The specific asymmetry  in  frequencies of the two branches, resulting in the oscillatory  behaviour of the under-damped branch only,  is due to  the special relationship  between ratios  (\ref{ratio}). It is intriguing that  the expressions are no more complex than those in the Kerr case \cite{Cvetic:2013}. 
In particular, eq.   (\ref{qmode}) agrees with  eq. (0.28) of \cite{Cvetic:2013}
which was obtained for the subtracted geometry  of the neutral Kerr solution, i.e.  the case with $\delta_i=0$, and thus $\Pi_c=1$ and $\Pi_s=0$.

The subtracted geometry has a remarkable property that in the near-BPS limit ($m\to 0$, $a\to 0$, $\delta_i\to \infty$, with $m e^{2\delta_i}$ and ${m}{a}^{-1}$ finite) the near-horizon geometry of such black holes and their subtracted geometry are the same. As a consequence, the quasi-normal modes of the near-BPS black holes  and those of  their subtracted geometry are the same\footnote{ We are grateful for Shahar Hod for pointing out to us after the appearance
of \cite{Cvetic:2013}  that if one specialises to the
near-BPS case of   slowly rotating ($a\ll m$) Kerr-Newman black holes then  $\beta_R \simeq 2  \beta _H$ and    the   family of modes given by eq. (11) of \cite{Hod:2008se} 
have identical frequencies to those of the second family of modes
in eq. (0.28) of \cite{Cvetic:2013} and hence to the second family
of (\ref{qmode}) of this paper.  The first family  of (\ref{qmode})  in this limit corresponds to negative imaginary frequencies  whose absolute values are much larger than those of the second family, and thus this  (ultra-damped) branch did not appear in \cite{Hod:2008se}.}.


\section{ Subtracted Magnetised Geometry}

The original subtracted Melvin metric was derived in 
\cite{Cvetic:2013roa} as a scaling limit of  magnetised STU black holes.   
It  describes  a generalization of the 
(static) subtracted geometry,  parameterised by  an additional magnetic 
field parameter $\beta_4$ which is  associated with the magnetic component of the Kaluza-Klein gauge 
field ${\cal A}_2$. The full solution is given in the Appendix 5.2.

Remarkably, one  may cast this metric in the same form as 
the rotating subtracted metric (\ref{metricn}), which we display again
\be
ds ^2 = \sqrt{\Delta} \frac{X}{F^2}
 \Bigl (- dt ^2 + \frac{F^2 dr^2} {X^2 } \Bigr )+
  \sqrt{\Delta} d\theta ^2 + \frac{F^2 \sin ^2 \theta}{ \sqrt{\Delta}  }  (d \phi + W d t) ^2  \, , \label{metricm}
\ee 
where now
\bea
X&=&r^2-2mr\, ,\nn\\
F^2&=&(2m)^3\left[(\Pi_c^2-\Pi_s^2)r +(2m)\Pi_s^2\right]\, ,\nn \\
W&=&-\frac{16m^4\Pi_s\Pi_c\beta_4}{F^2}\, , \nn\\
\Delta&=&F^2+(2m)^6\beta_4^2(\Pi_c^2-\Pi_s^2)^2\sin^2\theta\, . \nn\\
\eea
This is effectively  a generalization of the static subtracted geometry  with the magnetic field parameter $\beta_4$  introducing a specific spatial rotation. 
The metric has two horizons
\be r_+=2m\, , \quad r_-=0\, .
\ee
The inverse surface gravities of the inner and outer horizon are  
determined by
\be
\frac{1}{\kappa_+}=\frac{2F(r_+)}{r_+-r_-}=4m\Pi_c\, , \quad \frac{1}{\kappa_-}=\frac{2F(r_-)}{r_+-r_-}=4m\Pi_s\, , \label{sgm}\ee 
and  are the same as the inverse surface gravities for the static subtracted geometry, i.e.  (\ref{kappa}) with $a=0$. 
The angular velocities at the inner and outer horizon are are given by 
\be
\Omega_+=-W(r_+)=  \beta _4 \frac{\Pi _s }{\Pi _c} \, , \quad  \Omega_-=-W(r_-)=  \beta _4 \frac{\Pi _c }{\Pi _s} \,  .  \label{omm}
\ee
Note that in this case the ratios
\be
\frac{\Omega_+}{\kappa_+}=4m\beta_4\Pi_s\, , \quad \frac{\Omega_-}{\kappa_-}=4m\beta_4\Pi_c\, ,
\ee
are different, and  now the  radial part of the massless scalar wave equation   (\ref{wave})  
depends on  both  independent ratios.

\subsection{Kruskal-Szekeres  Coordinates for Subtracted Magnetised Geometry}

The retarded and advanced co-rotating  Eddington-Finkelstein  
coordinates are of the same form  as in (\ref{EF})
and the Killing vector $l^+$ (\ref{lp})  again coincides with the null generator on the horizon.

We introduce the Kruskal-Szekeres coordinates (\ref{KS})   which yield (\ref{UV}) and the metric (\ref{metricm}) takes the form (\ref{metricnn}).  In the vicinity of the outer horizon  $r\sim 2m$  one obtains  $-UV\sim  (r-2m)$,  and thus the metric (\ref{metricnn}) is regular and analytic there.
 
 We calculate the length squared of the co-rotating Killing vector $l^+$ (\ref{lp})
\bea
g^{\alpha \beta} l^+_\alpha l^+_\beta&=& - \frac{\sqrt{\Delta_s}}{F}\frac{(r-2m)}
{ (\Pi_c^2 -\Pi _s ^2 ) r + 2m \Pi _s^2 + 8 m^3 \beta _4 ^2 \sin^2\theta ( \Pi _c^2 -\Pi_s^2 )^2 } \nonumber \\
& \times&  \left[(\Pi_c^2 - \Pi_s^2) r + 2m  \Pi_s^2 
\right]\times \left[ r + 8m^3 \beta ^2_4\sin^2\theta \frac{1}{\Pi^2 _c}
( \Pi _c ^2 - \Pi _s ^2 ) ^2      \right]\,,    
\eea
which is negative outside the horizon, $r>2m$. Thus, this geometry is stable with  no super-radiance.

\subsection{Massless Wave Equation and Quasi-Normal Modes} 
The massless wave equation is again separable with the same wave function Ansatz as (\ref{ansatz}). The radial wave equation   can be cast in the same form as  (\ref{wave}) with the  inverse surface gravities  (\ref{sgm})  and angular velocities   (\ref{omm}). 

Solutions which are ingoing on the future horizon
must be regular at $U=0$ in Kruskal-Szekeres coordinates and
this implies that \cite{Cvetic:1997uw,Cvetic:1997xv,Cvetic:2011dn}   
\bea
&&\chi(x)= (x+\half)^{- (l+1) }
\bigl( \frac{x-\half}{x+\half } \bigr ) ^{ -i (\omega - n \Omega _+) \frac{\beta_H }{4 \pi }}\nn\\
&&F ( l+1-  i  \frac{\beta_R \omega -n (\beta _H \Omega _++\beta_-\Omega_-)}{4 \pi},
 l+1-  i  \frac{\beta_L\omega-n(\beta_H\Omega_+-\beta_-\Omega_-) }{4 \pi}, 1-   i \frac{\beta_H (\omega - 
n \Omega _+ ) }{2 \pi }
; \frac{x-\half}{x+\half } ) \,,  \nn\eea
where  again
\ben
\frac{\beta _H}{2 \pi} = \frac{1}{\kappa_+} \,, \quad \frac{\beta_-}{2\pi}=\frac{1}{\kappa_-}\, \quad 
 \frac{\beta _R}{2 \pi } = \frac{1}{\kappa_+} + \frac{1}{\kappa _-} \,,\quad 
\frac{\beta _L}{2 \pi } = \frac{1}{\kappa_+} - \frac{1}{\kappa _-}\,.  
\een
Near the outer horizon  $r^\star \rightarrow - \infty,\,
(x-\half)(x+\half) ^{-1} \rightarrow e^{2 \kappa_+ r^\star } 
$  and so 
\bea
 &&\chi(x) \approx 
e^{-i  (\omega - 
n \Omega _+)  r^\star} \nn\\
&&F( l+1-  i  \frac{\beta_R \omega -n (\beta _H \Omega _+ +\beta_-\Omega_-) }{4 \pi},
 l+1-  i  \frac{\beta_L\omega-n(\beta_H\Omega_+-\beta_-\Omega_-) }{4 \pi}, 1-   i \frac{ \beta_H(\omega  - n \Omega _+)  }{2 \pi }
; e^{2 \kappa _+ r^\star}  )\, .\nn
\label{ingoingm} \eea

In Kruskal-Szekeres coordinates therefore
\ben
e^{-i\omega t} e^{in \phi} \chi (x) \approx e^{in \phi_+ } 
 V^{-i\frac{\omega - n \Omega_+}{\kappa _+ } }  ( 1 + \dots )   
\een
where the ellipses denote a power series in $UV$
which is convergent in a neighbourhood of the future horizon 
$U=0$\,.  

At large $x$  \cite{Cvetic:1997uw,Cvetic:1997xv}
\bea
&&\chi(x) \approx  x ^{-(l+1)}
\frac{
\Gamma (1 -i  \frac{\beta_H (\omega -n  \Omega _+) }{2 \pi }) \Gamma (-2l-1 ) }
 {\Gamma (-l -i \frac{\beta_L \omega-n(\beta_H\Omega_+-\beta_-\Omega_-)}{4 \pi})
 \Gamma (-l -i  \frac{\omega \beta_R - n (\beta _H \Omega _++\beta_-\Omega_-) }{4 \pi}  )
}\nn\\
&&+ x^l  
\frac{\Gamma (1 -i  \frac{\omega \beta_H (\omega -n\Omega _+)  
}{2 \pi }) \Gamma (2l  +1 )} 
{\Gamma ( l+1 - i  \frac{\beta_L\omega-n(\beta_H\Omega_+-\beta_-\Omega_-)}{4 \pi}) 
\Gamma (l+1 -i   \frac{\omega \beta_R - n (\beta _H \Omega _++\beta_-\Omega_-) } {4 \pi}) 
} \, . 
\eea
In order that $\chi$ be finite at spatial infinity, we must set  
\bea
&&i  \left( \frac{\omega \beta_L}{4\pi}-n\frac{\beta_H\Omega_+-\beta_-\Omega_-}{4\pi}\right) = l+1 +N_L
\,, \nn \\
\hbox{or}\quad
&&i    \left(\frac{\omega \beta_R}{4\pi} - n\frac{ \beta _H \Omega _++\beta_-\Omega_-}{4 \pi}\right)
=l+1 +N_R \,, 
\eea
where $N_{L,R}=0,1,\dots$
This gives remarkably simple  and symmetric  formulae for the frequencies of the quasi-normal modes
\bea
&&\omega=-\frac{i}{2m(\Pi_c-\Pi_s)}(1+l+N_L)-n\beta_4\, , \nn\\
\hbox{or}\quad &&\omega=-\frac{i}{2m(\Pi_c+\Pi_s)}(1+l+N_R)+ n\beta_4\,.
\eea
Both  frequencies result in damped modes with  a symmetric shift in  advanced and retarded oscillatory behaviour due to the magnetic field parameter $\beta_4$.

An interesting observation can be made here about the magnetic field parameter in the above quasi-normal modes. According to the Bohr's correspondence principle, the frequency of oscillation of a classical system is equivalent to the frequency of transition of the corresponding quantum system. Guided by this principle, in \cite{Hod},  some observations were made which indicate that the real part of the quasi-normal modes is related to the quantized area spectrum of the  quantum black hole. In our case the real part of the quasi-normal modes is related in a very simple way to the magnetic field parameter, thus making it easy to see how turning on the magnetic field affects the area spectrum of the quantum black hole.

\section{Lifted Geometries and Quasi-Normal Modes}

In Appendix 5.3  we derive  the explicit lift of the subtracted geometries on a circle of size $2\pi R$ and parameterised by a coordinate $z$. The five-dimensional geometry is locally ${\rm BTZ} \times S^2$  with  the BTZ coordinates  denoted by $\{t_3,r_3,\phi_3\}$ and  the $S^2$ coordinates denoted by $\{\theta,{\bar \phi}\}$.  The explicit  transformation  between $\{t,r,\theta,\phi,z\}$ coordinates,  and the ${\rm  BTZ} \times S^2$  coordinates  is given in the Appendix 5.3, too. The BTZ metric (\ref{BTZ}) can also be cast into local AdS$_3$ metric (\ref{ads3m}), parameterised by  coordinates $\{T,\rho,\Phi\}$. The explicit transformation between the BTZ  and the local AdS$_3$ coordinates is given in Appendix 5.4, following \cite{Banados:1992wn,Banados:1992gq}. The radius of AdS$_3$ is $\ell$ and the radius  of $S^2$  is $\frac{\ell}{2}$.  Specifically, $\ell=4m(\Pi_c^2-\Pi_s^2)^{\frac{1}{3}}$.  

Since for this five-dimensional geometry the wave equation for the minimally coupled massive  scalar field is separable  and exactly solvable, this allows us to study  explicitly the quasi-normal modes directly in five dimensions.   Furthermore, the scalar field wave function can be expanded in terms of Kaluza-Klein modes, parameterised by a quantised wave number $k$ along  the circle   direction $z$. We can therefore  study the quasi-normal modes for each Kaluza-Klein mode by  solving directly the wave equation in five dimensions  for the  complete tower of  Kaluza-Klein  states, i.e. we do not have to  resort to solving a complicated  equation for each Kaluza-Klein  mode separately.

The wave equation  for a massive,  minimally coupled scalar field $\Phi$ in the local AdS$_3\times S^2$  background  is separable and solved  with the Ansatz
\be 
{\bf \Phi}=  e^{-i{\bar \omega}T}e^{i{\bar k}\Phi}e^{ i { n}{\bar \phi}}\, P^{ n}_l ( \cos \theta)\, \chi(\rho) \, .\label{an5}
\ee
$ P^n_l (\cos\theta)$, the associated Legendre function, is a solution for the Laplacian of  the unit two-sphere $S^2$ with eigenvalues $l(l+1)$. Here $n = 0 , \pm1 ,\pm 2 ...\pm l$ and $l$ is a non-negative integer.
Again,  $\{T,\Phi,\rho\}$  and $\{\theta,{\bar \phi}\}$ parameterise the local AdS$_3$ and  $S^2$ coordinates, respectively.
Furthermore, in our context  the radius of AdS$_3$  is $\ell$ and that of $S^2$ is $\frac{\ell}{2}$ where we have $\ell=4m(\Pi_c^2-\Pi_s^2)^{\frac{1}{3}}$ (see Appendix 5.3).

The metric, describing a local AdS$_3$  (\ref{ads3m}) 
 \be
d s_{AdS_3}^2 = \ell^2\, 
  (-\sinh^2\rho\, dT^2 + d\rho^2 + \cosh^2\rho\, d\Phi^2)\label{adsl}
\,.
\ee
has the Laplacian
\begin{equation}
\Box_{AdS_3} = \partial_\rho^2 + \frac{2\cosh(2\rho)}{  \sinh(2\rho)} \partial_\rho
- \frac{1}{  \sinh^2\rho} \partial_T^2 + \frac{1}{ \cosh^2\rho}
\partial_\Phi^2\, ,
\end{equation}
and enters the  five-dimensional Klein-Gordon equation equation in the following form:
 \begin{equation}
[\ell^2\left(\Box_{AdS_3}- 4l(l+1)\right)-M_5^2 ]{\bf \Phi}=0
\end{equation}
Note again  that  $4\ell^2l(l+1)$ is the eigenvalue of the two-sphere $S^2$ Laplacian with the two-sphere radius $\frac{\ell}{2}$. For the Ansatz (\ref{an5}) this equation becomes
\be
\Bigl[\ell^2\bigl( \partial_\rho^2 +\frac {2\cosh(2\rho)}{ \sinh(2\rho)} \partial_\rho
+\frac{{\bar \omega}^2}{ \sinh^2\rho} - \frac{{\bar k}^2}{  \cosh^2\rho}
-4l(l+1)\bigr)-M_5^2\Bigr])\chi(\rho)=0\,.\label{radial}
\ee
The solution, corresponding to the incoming wave at the outer horizon, is 
\bea
&&\chi(\rho)= (x+\half)^{- ({\bar l}+1) }
\bigl( \frac{x-\half}{x+\half } \bigr ) ^{ -i \frac{\bar \omega}{2}}\nn\\
&&F ({\bar  l}+1-  i  \frac{({\bar\omega}+{\bar k})}{2},
{\bar  l}+1-  i  \frac{({\bar\omega}-{\bar k})}{2}
, 1-   i {\bar \omega}
; \tanh^2\rho) \, . \label{hgs} \eea
Here we have introduced
\be 
{\bar l}({\bar l}+1)\equiv l(l+1)+\frac{M_5^2}{4\ell^2}\, .
\ee
While the analysis can be completed  for massive minimally coupled five-dimensional scalars, in the following we will focus on massless ones, i.e. taking $M_5=0$ and thus ${\bar l}=l$. The only quantitative difference in the analysis for massive five-dimensional scalars is that the expressions below involve a change $l\to {\bar l}> l$, and  thus a shift in the quasi-normal  frequencies.

At this point we  relate the respective  local AdS$_3$ and $S^2$ coordinates 
$\{T,\Phi,\rho\}$ and $\{\theta,{\bar \phi}\}$  
to $\{t,r,\,\theta,\phi,z\}$. This can be done by first employing Appendix 5.3, where the explicit lift to the  ${\rm BTZ} \times S^2$   and the map to the BTZ  and   $S^2$ coordinates is given,
and then  employing  Appendix 5.4, where   the transformation between the  
BTZ   and  local AdS$_3$ coordinates is provided. The result for the subtracted rotating geometry is
\bea
T&=&\frac{4\sqrt{m^2-a^2}}{\ell^3}(\frac{t}{\kappa_+}-\frac{z}{\kappa_-})\, , \nn\\
\Phi&=&\frac{4\sqrt{m^2-a^2}}{\ell^3}(\frac{z}{\kappa_+}-\frac{t}{\kappa_-})\, , \label{TPhi}
\eea
and
\be
\cosh^2\rho=x+\frac{1}{2}\, , \quad \sinh^2\rho=x-\frac{1}{2}\, ,\label{radialr}
\ee
where $x$ is defined in (\ref{dlr}), i.e. $x=\left[r-\frac{1}{2}(r_++r_-)\right](r_+-r_-)^{-1}$.
Furthermore, for $S^2$ coordinates, $\theta$ is unchanged and the azimuthal angle $\bar \phi$ is related to $\phi$ as in  (\ref{barp}):
\be  {\bar \phi}=\phi -\frac{16ma(\Pi_c-\Pi_s)}{\ell^3}(z+t)\, . \label{bphi}
\ee
The  $2\pi$ periodicity  of ${\bar \phi}$ is ensured if ${16ma(\Pi_c-\Pi_s)}{\ell^{-3}}
=a(2m)^{-2}(\Pi_c+\Pi_s)^{-1}
$ is quantized in units of $R^{-1}$.

The radial equation   (\ref{radial})  can be cast in the following form:
\be
\bigl[ \partial_x(x^2-\frac{1}{4})\partial_x+{{\bar \omega}^2 \over{ 4(x-\frac{1}{2})}} \ - {{\bar k}^2  \over{4(x+\frac{1}{2})}} 
-l(l+1)
\bigr]\chi(x)=0\, .\label{radialpp}
\ee
The above  coordinate transformations allow us to  relate the quantum numbers in the Ansatz (\ref{an5}) to those of the standard Kaluza-Klein Ansatz:\footnote{By abuse of notation we use above the same  radial function notation.}
\be
{\bf \Phi}=  e^{-i{\omega}t}e^{i{ k}z }e^{ i { n}{{\phi}}}\, P^n_l ( \cos \theta)\,\chi(r)  \, .\label{ankk}
\ee
 Namely, equating the two Ans\"atze (\ref{an5}) and (\ref{ankk}),  and  employing the coordinate transformations  (\ref{TPhi}) and  (\ref{bphi}) yields the following transformation between  quantum numbers  $\{{\bar \omega},{\bar k} \}$ and $\{\omega,k\}$:
\be
{\bar \omega}=\frac{\omega}{\kappa_+}-\frac{k}{\kappa_-}-n\frac{\Omega_+}{\kappa_+}\, , \quad
{\bar k}=-\frac{\omega}{\kappa_-}+\frac{k}{\kappa_+}+n\frac{\Omega_+}{\kappa_+}\, , \label{master}
\ee
and $n$ unchanged.

For the subtracted magnetised  geometry the expressions  for  (\ref{TPhi}) are the same, but with  $a=0$  and  static expressions for inverse surface gravities (\ref{sgm}), i.e.  ${\kappa_+^{-1}}=4m\Pi_c$ and $\kappa_-^{-1}=4m\Pi_s$.  The azimuthal angle is shifted due to the magnetic field $\beta_4$ as in (\ref{barpm}):
\be {\bar \phi}=\phi -\beta_4 z\, .
\ee
Note that $2\pi$ periodicity of the  $S^2$ azimuthal angle ${\bar \phi}$ is ensured if
the magnetic field parameter $\beta_4$ is  quantised in units of $ R^{-1}$.

As a consequence,  the transformation between the quantum numbers $\{{\bar\omega},{\bar k}\}$  and $\{\omega,k\}$ is
 \be
{\bar \omega}=\frac{\omega}{\kappa_+}-\frac{k+n\beta_4}{\kappa_-}\, , \quad
{\bar k}=-\frac{\omega}{\kappa_-}+\frac{k+n\beta_4}{\kappa_+}\, , \label{masterm}
\ee
and again, $n$ unchanged.

These  general expressions now allow us to recover  results for the massless four-dimensional field with vanishing wave number $k=0$. For the   subtracted rotating  geometry one obtains
\be
{\bar \omega}=\frac{\omega}{\kappa_+}-n\frac{\Omega_+}{\kappa_+}\, , \quad {\bar k}=-\frac{\omega}{\kappa_-}+n\frac{\Omega_+}{\kappa_+}\, , \label{masterk}
\ee
just as in Section 2. Similarly for the magnetised subtracted geometry:
\be
{\bar \omega}=\frac{\omega}{\kappa_+}-\frac{n\beta_4}{\kappa_-}\, , \quad
{\bar k}=-\frac{\omega}{\kappa_-}+\frac{n\beta_4}{\kappa_+}\, , \label{mastermk}
\ee
in agreement with Section 3.

We can also study massive Kaluza-Klein modes with  the wave number $k\ne 0$, which is quantised in units of $R^{-1}$, where $R$ is the radius of the circle $S^1$.
Those are massive  four-dimensional particles with  mass  $m_4\propto k$, and they are charged  under the Kaluza-Klein  U(1)  gauge  symmetry  with the  charge $k=q$  (see Appendix 5.5).   Their quasi-normal modes can be determined completely analogously to massless modes in Sections 2 and 3.

The solution  (\ref{hgs}), corresponding to the incoming wave at the outer horizon,  is  required to be finite at a large $x$, which is achieved  for
\be
\frac{{\bar \omega}+{\bar k}}{2}=-i(1+l+N_L)\, , \quad \hbox{or}\quad \frac{{\bar \omega}- {\bar k}}{2}=-i(1+l+N_R)\, , 
\ee
where $l=0,1,\dots$, and $N_L=0,1,\dots$ or $N_R=0,1,\dots$  This constrains a specific combination of  $\omega$ and $k$. In the rotating case we have
\bea
&&\omega=-\frac{i}{2m(\Pi_c-\Pi_s)}(1+l+N_L)+k\, , \nn\\
\hbox{or}\quad &&\omega=-\frac{i\sqrt{m^2-a^2}}{2m^2(\Pi_c+\Pi_s)}(1+l+N_R)+ \frac{a}{2m^2(\Pi_c+\Pi_s)}n-k\,.
\eea
In the  subtracted magnetised  case we obtain
\bea 
\omega&=&-\frac{i}{2m(\Pi_c-\Pi_s)}(1+l+N_L)+n\beta_4+k\, ,\nn\\
\hbox{or}\quad \omega&=&-\frac{i}{2m(\Pi_c+\Pi_s)}(1+l+N_R)-n\beta_4-k\,. \
\eea
Again, we obtained two branches of damped quasi-normal modes,  both with oscillatory behaviour symmetrically advanced and retarded by  $n\beta_4+k$. 
 
It is interesting to point out that the solution (\ref{hgs}) for massive modes with $k\ne 0$ has a regular, analytic  behaviour near the  outer horizon, after one has made a gauge transformation $\chi(x)\to   e^{ik{\cal A}_{2t+}t }  \chi(x)$, where ${\cal A}_{2t+}=(2m)^4\Pi_c\Pi_sF^{-2}(r_+)$ is the time component of the Kaluza-Klein gauge potential  ${\cal A}_2$ (\ref{gaugep}) or (\ref{gpm}), evaluated at the outer horizon $r_+$. Namely, we obtain
\bea
e^{ik{\cal A}_{2t+}t}e^{-i\omega t} e^{in \phi} \chi (x)&\approx& e^{ik{\cal A}_{2t+}t}e^{-i(\omega-n\Omega_+) t} e^{in {\phi_+}} e^{-i{{\bar \omega}\kappa_+r*}} (1+\cdots)\nn\\
&\approx& e^{in \phi_+ } 
 V^{-i\frac{\omega - n \Omega_+}{\kappa _+ } +i\frac{k}{\kappa_-}}  ( 1 + \dots )   \, ,
\eea
where  we wrote the final expression in terms of Kruskal-Szekeres coordinates,  and the ellipses denote a power series in $UV$
which is convergent in a neighbourhood of the future horizon 
$U=0$ \,. 

\vskip 1cm

 \noindent{\bf Acknowledgements} We would like to thank  Finn Larsen and Monica Guica for useful discussions.
 This research is supported in part by the DOE grant DE-SC0007901,  the Fay R. and Eugene L. Langberg Endowed
Chair (M.C.) and the Slovenian Research Agency (ARRS) (M.C.).

\newpage

\section{Appendices}
\subsection{Subtracted Rotating   Geometry  with Sources}
In \cite{Cvetic:2012tr} it was shown that  the subtracted geometry  (\ref{metric4d}), (\ref{c4d}), (\ref{warpf}) for  four-charge rotating black hole is a solution of the equations of motion for the  STU Lagrangian, describing the bosonic part of the  N=2 supergravity Lagrangian coupled to three vector super-multiplets:

\bea
{\cal L}_4 &=& R\, {*{\bf 1}} - \frac{1}{2} {*d\varphi_i}\wedge d\varphi_i 
   - \frac{1}{2} e^{2\varphi_i}\, {*d\chi_i}\wedge d\chi_i - \frac{1}{2} e^{-\varphi_1}\,
( e^{\varphi_2-\varphi_3}\, {*  F_{1}}\wedge   F_{ 1}\nn\cr
 &+& e^{\varphi_2+\varphi_3}\, {*   F_{ 2}}\wedge   F_{ 2}
  + e^{-\varphi_2 + \varphi_3}\, {*  {\cal F}_1 }\wedge   {\cal F}_1 + 
     e^{-\varphi_2 -\varphi_3}\, {* {\cal F}_2}\wedge   {\cal F}_2)\nn\\
&-& \chi_1\, (  F_{1}\wedge  {\cal F}_1 + 
                   F_{ 2}\wedge  {\cal F}_2)\,,
\label{d4lag}
\eea
where the index $i$ labelling the dilatons $\varphi_i$ and axions $\chi_i$
ranges over $1\le i \le 3$.  The four U(1)  field strengths can be written in 
terms of potentials as
\bea
  F_{ 1} &=& d   A_{1} - \chi_2\, d {\cal A}_2\,,\nn\cr
  F_{ 2} &=& d  A_{ 2} + \chi_2\, d {\cal A}_1 - 
    \chi_3\, d   A_{ 1} +
      \chi_2\, \chi_3\, d  {\cal A}_2\,,\nn\cr
  {\cal F}_1 &=& d  {\cal A}_1 + \chi_3\, d  {\cal A}_2\,,\nn\cr
  {\cal F}_2 &=& d  {\cal A}_2\,.
\eea

 The  three axio-scalar fields  and the four U(1) gauge potentials  can be formally obtained as a scaling limit  of a  certain  black hole solution (for details, see \cite{Cvetic:2012tr}), resulting in
\be
\chi_1=-\chi_2=\chi_3=-\frac{2ma(\Pi_c-\Pi_s)\cos\theta}{Q^2} , \ \ e^{\varphi_1} =e^{\varphi_2} =e^{\varphi_3} = \frac{Q^2} {\sqrt{\Delta}}\, ,\label{scalars}
\ee
and  the  gauge potentials $A_1=A_2=A_3\equiv A $ for gauge field strengths    $* F_1= F_2= * {\cal F} _1\equiv F\, $  and   ${\cal A}_2$ for  $ {\cal F}_2$ are of the following form:
\bea
A=&& -\frac{r-m}{Q}\, dt 
  -\frac{(2\, m)^2 \, a^2\, (\Pi_c-\Pi_s)[\,r\,(\Pi_c-\Pi_s)+2m\Pi_s\,]\, \cos^2\theta}{Q\Delta}\, dt
\cr
&&
-\frac{(2m)^4\, a\, (\Pi_c-\Pi_s)[\, r\, (\Pi_c^2-\Pi_s^2)+2m\Pi_s^2\, ]\, \sin^2\theta}{Q\Delta}\, d\phi\,  ,\label{orgp}\\
{\cal A}_2=&& \frac{ Q^3[(2m)^2 \Pi_c\Pi_s + a^2 (\Pi_c-\Pi_s)^2\cos^2\theta]}{2m(\Pi_c^2-\Pi_s^2)\Delta}\, dt \, + \frac{Q^32m\,a(\Pi_c-\Pi_s)\sin^2\theta}{\Delta}   \, d\phi\,, 
\label{gaugep}
\eea

where
\be
Q = {2m} (\Pi_c^2-\Pi_s^2)^{\frac{1}{3}}{\epsilon^{-\frac{1}{3}}}\equiv  \textstyle{\frac{1}{2}}{\ell}\epsilon^{-{\frac{1}{3}}}, \quad \hbox{as}\quad  \epsilon \to  0 
\, .\label{Q}
\ee
and again, $\Delta$ defined  as  in (\ref{warpf}):
\be
\Delta_0 \rightarrow  \Delta =  (2m)^3r(\Pi_c^2-\Pi^2_s) +(2m)^4\Pi_s^2
-(2m)^2 (\Pi_c-\Pi_s)^2 a^2 \cos^2 \theta \,. \ee
 The  (formally infinite) factors of Q can in principle  be removed from gauge potentials  by removing   corresponding factors from scalar fields. 
 However, when lifting the scaling limit solution to five dimensions,  it is useful to keep this scaling factor explicit; in the final five-dimensional metric an overall factor is  not relevant.

 \subsection{Subtracted Magnetised Geometry with Sources}

The  magnetised solution of the static STU black hole was obtained in \cite{Cvetic:2013roa} and is of the form:
\be
ds_4^2 = H\, [- r(r-2m) dt^2 + \fft{r_1 r_2 r_3 r_4}{r(r-2m)}\,dr^2 +
   r_1 r_2 r_3 r_4 d\theta^2\,]  
   + H^{-1}\, \sin^2\theta\, (d\phi -{\tilde \omega} dt)^2\,.
\label{4metric}
\ee
Here
\be
r_i= r + 2m s_i^2\,,
\ee
and we shall use the  notation $s_i=\sinh\delta_i$ and $c_i=\cosh\delta_i$, with $i=1,2,3,4$.
The function ${\tilde \omega}$ is given by
\be
{\tilde \omega} = \sum_{i=1}^4 \Big[ -\fft{q_i\, \beta_i}{r_i} +
 \fft{q_i\,\Xi_i\,  [r_i+(r-2m)\cos^2\theta] r}{r_i}\Big]\,,
\label{omega}
\ee
where
\be
q_i= 2 m s_i c_i\,,\qquad 
\Xi_i= \fft{\beta_1\beta_2\beta_3\beta_4}{\beta_i}\,,\qquad 
\beta_i=\ft12 B_i\,,
\ee
and $B_i$ ($i=1,2,3,4$) denote the external magnetic field strengths for each of the 
four gauge fields.  Finally, the function $H$ is given  by
\be
H= \fft{\sqrt{{\bar \Delta}}}{\sqrt{r_1 r_2 r_3 r_4}}\,,\label{Hdef}
\ee
where
\bea
{\bar \Delta} &=& 1 + \sum_i\fft{\beta_i^2 r_1 r_2 r_3 r_4}{r_i^2}\, \sin^2\theta +
  2[\beta_3\beta_4 q_1 q_2+\cdots]\cos^2\theta
+[\beta_3^2\,\beta_4^2 \,R_1^2\, R_2^2 +\cdots]\nn\\
&& - 2 (\prod_j \beta_j r_j)\,  \sum_i\fft{q_i^2}{r_i^2}\, 
   \sin^2\theta\cos^2\theta 
 +[2\beta_2\beta_3\beta_4^2 q_2 q_3 \, R_1^2
   +\cdots]\cos^2\theta  +\prod_i \beta_i^2 \,R_i^2 \nn\\
&& + r_1 r_2 r_3 r_4\, 
\sum_i \fft{\Xi_i^2 \, R_i^2}{
    r_i^2}\, \sin^2\theta 
+
 [2\beta_1\beta_2\beta_3^2\beta_4^2 q_3 q_4\,
 R_1^2 \,R_2^2
  +\cdots]\cos^2\theta\,,
\eea
and we have defined 
\be
R_i^2 = r_i^2 \, \sin^2\theta + q_i^2\, \cos^2\theta\,.
\ee
The Kaluza-Klein gauge field here is given by
\be
{\cal A}_2 = \Big[\fft{q_4}{r_4} - \sum_{i=1}^3 
  \fft{r\, q_i\, \beta_1\beta_2\beta_3\, [r_i+ (r-2m)\cos^2\theta]}{
            \beta_i\, r_i}\Big]\, dt - \sigma_4\, (d\phi-{\tilde\omega} dt)\, , 
 \ee                  
where $ \sigma _4={\tilde{\sigma_4}} {\bar \Delta}^{-1}$, and 

\bea
\tilde\sigma_4 &=& \fft{\beta_4 r_1 r_2 r_3}{r_4}\sin^2\theta +
  (\beta_1 q_2 q_3+\cdots) \cos^2\theta 
+\beta_4(\beta_1^2 R_2^2 R_3^2+\cdots) \nn\\
&&+2\beta_4(\beta_2\beta_3 q_2 q_3 R_1^2+\cdots)\cos^2\theta
+ q_4[\beta_1^2(\beta_2 q_2 R_3^2 +\beta_3 q_3 R_2^2)+\cdots]\cos^2\theta\nn\\
&&+ 4\beta_1 \beta_2 \beta_3 q_1q_2 q_3 q_4 \cos^4\theta  \nn\\
&&-
 \fft{\beta_1\beta_2\beta_3 q_4^2 r_1 r_2 r_3}{r_4}\sin^2\theta\cos^2\theta
-\beta_1\beta_2\beta_3\,r_4\,
\Big(\fft{q_1^2 r_2 r_3}{r_1} +\cdots\Big)\sin^2\theta\cos^2\theta \nn\\
&&
+\beta_1 \beta_2\beta_3(\beta_2\beta_3 q_2 q_3 R_1^2+\cdots) R_4^2\cos^2\theta 
 +\beta_4 r_4\Big[\fft{\beta_2^2 \beta_3^2 r_2 r_3}{r_1}\, R_1^4+\cdots\Big] 
\sin^2\theta \nn\\
&&+ 2\beta_1\beta_2\beta_3\beta_4 q_4(\beta_1 q_1 R_2^2 R_3^2 +
\cdots)\cos^2\theta + \beta_4\beta_1^2 \beta_2^2\beta_3^2 R_1^2 R_2^2 
R_3^2 R_4^2\,.
\eea            
The dilaton field is given by 
\be
e^{\varphi_1} =\fft{Y_1}{\sqrt{{\bar \Delta}\, r_1 r_2 r_3 r_4}}\,,\
\ee
where
\bea
Y_1 &=& r_1 r_3(1+2 \beta_1\beta_3 q_2 q_4 \cos^2\theta +
\beta_1^2 \beta_3^2 R_2^2 R_4^2) \nn\\
&+& r_2 r_4(\beta_1^2 R_3^2+\beta_3^2 R_1^2 
  + 2\beta_1\beta_3 q_1 q_3 \cos^2\theta)\,.
\eea
For explicit expressions of all the  fields  see \cite{Cvetic:2013roa}. Note however in order to have the same sign for the gauge fields of the rotating and magnetised geometries, we have changed an overall sign for the gauge fields relative to \cite{Cvetic:2013roa}.

\vskip0.2cm
\noindent{\bf The Scaling Limit}
\vskip 0.2cm
{\noindent   The subtracted geometry can be obtained by taking a scaling limit of the above
magnetised electric black holes,  analogously to  the rotating case.}
The limit can be implemented by means of the scalings 
\bea
&&m\to m\, \ep\,,\qquad r= r\, \ep\,,\qquad
    t\to  t\,\ep^{-1}\,,\qquad \beta_i \to \beta_i\, \ep\,,\quad 
i=1,2,3,4\,,\nn\\
&&\sinh^2\delta_4\to  \fft{\Pi_s^2}{\Pi_c^2-\Pi_s^2}\,,\qquad
\sinh^2\delta_i\to  (\Pi_c^2-\Pi_s^2)^{\frac{1}{3}}\, \ep^{-\frac{4}{3}}\,,\quad
i=1,2,3\,,\label{scaling}
\eea
where $\ep$ is then sent to zero. In particular, this gives
\be
(d\phi-{\tilde \omega} dt) \longrightarrow d\phi -
   (\beta_1+\beta_2+\beta_3) d t 
  -\fft{2 m \beta_4\, \Pi_c \Pi_s}{
    (\Pi_c^2-\Pi_s^2)  r + 2  m \Pi_s^2}\, d t\,,
\ee
and
\be
{\bar \Delta} \longrightarrow 1 + \fft{(2 m)^3 \beta_4^2 
   (\Pi_c^2-\Pi_s^2)^2 \sin^2\theta}{
          (\Pi_c^2-\Pi_s^2)  r + 2 m \Pi_s^2}\,,\quad r_1r_2r_3r_4\longrightarrow  (2m)^3\left[(\Pi_c^2-\Pi_s^2)r +2 m \Pi_s^2\right]\, .
\ee
The quantities $\beta_1$, $\beta_2$ and $ \beta_3$   are removed  by a gauge transformation
 $\phi\longrightarrow 
  \phi+(\beta_1+\beta_2+\beta_3)  t$.  We shall
assume from now on that this transformation has been performed. 
The final  metric can be cast in the following form:
\be
ds ^2 = \sqrt{\Delta} \frac{X}{F^2}
 \Bigl (- dt ^2 + \frac{F^2 dr^2} {X^2 } \Bigr )+
  \sqrt{\Delta} d\theta ^2 + \frac{F^2 \sin ^2 \theta}{ \sqrt{\Delta}  }  (d \phi + W d t) ^2  \, , \label{metricmm}
\ee 
where
\bea
X&=&r^2-2mr\, ,\nn\\
F^2&=&(2m)^3\left[(\Pi_c^2-\Pi_s^2)r +(2m)\Pi_s^2\right]\, ,\nn \\
W&=&-\frac{16m^4\Pi_s\Pi_c\beta_4}{F^2}\, , \nn\\
\Delta&=&F^2+(2m)^6\beta_4^2(\Pi_c^2-\Pi_s^2)^2\sin^2\theta\, . 
\eea
The  dilation fields are of the form:
\be
e^{\phi_1}  =e^{\varphi_2}= e^{\varphi_3}= \frac{Q^2}{ \sqrt{\Delta}}\, ,
\ee
and the axion fields vanish. The Kaluza-Klein U(1) gauge field becomes
\be
{\cal A}_2 = 
\frac{Q^3 2{ m}  \Pi_c \Pi_s }{ (\Pi_c^2-\Pi_s^2)F^2}dt-\frac{Q^3 (2m)^3 \beta_4( \Pi_c^2 -\Pi_s^2) \sin^2 \theta }{ \Delta}( d\phi+Wdt)
\, . \label{gpm}
\ee
Note that at the horizon the combination  $\phi+W(r_+)t=\phi_+$, and thus the second  term in (\ref{gpm})  becomes  the $\phi_+$ component  of the Kaluza-Klein gauge potential.  The remaining three gauge potentials  become identified and  are of the form (\ref{orgp}) by setting $a=0$.

One can of course remove $Q$  in the scalar and gauge fields via a  gauge transformation.  However, it is useful to keep it in the discussion of the lift and at the end remove the overall scaling parameter $\epsilon$.
 
\subsection{Subtracted Geometry Lifted to  Five Dimensions}

We now provide a lift of the  subtracted  rotating  geometry  to five-dimensions\footnote{Partial results were provided in \cite{Cvetic:2011dn,Cvetic:2012tr}. Here we take particular care  of the dimensions  
and  of  the periodicities of metric coordinates.}.  The five-dimensional metric for the scaling limit takes the form:

\be
ds_5^2 = e^{\varphi_1} ds_4^2\ + e^{-2 \varphi_1} ( dz + {\cal A}_2)^2\, ,
\ee
where we have to implement the scaling  $z\to z\epsilon^{-1}$.  This metric  takes the form:
\be
ds_5^2= \epsilon^{-\frac{2}{3}} (ds_{S^2}^2+ds_{BTZ}^2)\, ,   \label{5dlift}
\ee
where 
\be
ds_{S^2}^2=\textstyle{1\over 4}\ell^2\left(d\theta^2+\sin^2\theta d{\bar \phi}^2\right)\, ,  \label{S2}
\ee
with
\be
{\bar \phi}=\phi -\frac{16ma(\Pi_c-\Pi_s)}{\ell^3}(z+t)\, , \label{barp}
\ee
and
\be
ds_{BTZ}^2= -\frac{(r_3^2-r_{3+}^2)(r_3^2-r_{3-}^2)}{\ell^2\, r_3^2}
dt_3^2 + \frac{\ell^2r_3^2}{(r_3^2-r_{3+}^2)(r_3^2-r_{3-}^2)}\, dr_3^2+r_3^2(d\phi_3+\frac{r_{3+}r_{3-}}{\ell r_3^2}\, dt_3)^2\, ,
\label{BTZ}
\ee
where
\bea
\phi_3&=&\frac{z}{R}\, ,  \nn\\
t_3&=&\frac{\ell}{R} \, t\, ,\nn \\ 
 r_3^2&=&\frac{16(2mR)^2}{\ell^4}\left[2m(\Pi_c^2-\Pi_s^2)r+ (2m)^2\Pi_s^2-a^2(\Pi_c-\Pi_s)^2\right]\, . \label{BTZc}
\eea
Here, $R$ is the radius of the circle $S^1$ and $\ell= 4m(\Pi_c^2-\Pi_s^2)^{\frac{1}{3}}$ is  the radius of the AdS$_3$. 
Furthermore
\be
r_{3\pm}= \frac{8mR}{\ell^2} \left [m ( \Pi_c + \Pi_s) \pm \sqrt{ m^2 - a^2} (\Pi_c - \Pi_s)\right]\, .\label{r3pm}
\end{equation}
The periodicity of  $z$ coordinate is  $2\pi R$,  and thus  the angular coordinate $\phi_3$ has the correct periodicity of $2\pi$. 
Note also that the  $2\pi$ periodicity  of ${\bar \phi}$ is ensured if ${16ma(\Pi_c-\Pi_s)}{\ell^{-3}}
=a(2m)^{-2}(\Pi_c+\Pi_s)^{-1}
$ is quantized in units of $R^{-1}$.

The lifted geometry is indeed locally AdS$_3\times S^2$ with the radius of AdS$_3$
equal to  $\ell$ and the radius of $S^2$  equal to $\frac{\ell}{2}$.

\vskip 0.3cm
{\noindent{ \bf Subtracted Magnetised Geometry}  }
\vskip 0.3cm
{\noindent This geometry also lifts to (\ref{5dlift}) where now ${\bar \phi}$  in (\ref{S2}) is defined as\footnote{It was observed in \cite{Yaz} that such a shift produces a magnetic field for the Kaluza-Klein U(1) gauge potential and thus a four-dimensional geometry in a Kaluza-Klein magnetic field.}}
\be
{\bar \phi}= \phi -\beta_4\,  z\,,\label{barpm}
\ee
and we set in all expressions  above $a=0$, i.e.
  the  BTZ coordinates are related to  $\{t,r,z\}$  as in (\ref{BTZc}) with  $a=0$. 
(Obviously, $\beta_4=0$ corresponds to the lift of the static subtracted geometry.)
Note that the shift requires that  $\beta_4$ be quantized in units of  $R^{-1}$, in order for ${\bar\phi}$ to have the correct periodicity of $2\pi$.

\subsection{Relation of the BTZ Black Hole Coordinates to the  AdS$_{\bf  3}$ Coordinates}

According to  \cite{Banados:1992wn,Banados:1992gq} 
AdS$_3 $ is the quadric 
\ben
u^2 + v^2 -x^2 -y^2 = {\ell}^2 \, , 
\een
in ${\mathbb E} ^{2,2}$ 
with  the metric induced from 
\ben
ds ^2 = -du^2 -dv ^2 + dx^2 + dy ^2 \, . 
\een
In a local patch we have the embedding 
\bea
u &=&\sqrt{A (r)}  \cosh \Phi = {\ell}\cosh  \rho \cosh \Phi\, , \\
x &=& \sqrt {A (r)}  \sinh \Phi  ={\ell}\cosh  \rho  \sinh \Phi\, ,\\
y &=&\sqrt{B(r)}  \cosh T  =  {\ell}\sinh \rho \cosh T \, ,  \\ 
v &=& \sqrt{B(r)} \sinh T =  {\ell}\sinh \rho \sinh T\, .
\eea
The metric  is of the form:
\be
d s_{AdS_3}^2 = \ell^2\, 
  (-\sinh^2\rho\, dT^2 + d\rho^2 + \cosh^2\rho\, d\Phi^2)\label{ads3m}
\,.
\ee
The relationship to the BTZ metric coordinates and parameters introduced in the Appendix 5.3  (eqs.(\ref{BTZ},\ref{BTZc})) is
\ben
A(r) = {\ell}^2  \frac{r_3^2 - r_{3 -} ^2 }{r_{3+}^2 - r_{3-} ^2 } \, ,  \qquad  
B(r)= {\ell}^2 \frac{r_3^2 - r_{3+}^2 }{r_{3+}^2 - r_{3-}^2 }  \, ,  
\een
\ben 
T = \frac {r_{3+} t_3 -  r_{3 -}{\ell} \phi_3 }{{\ell}^2}  \qquad  
\Phi = \frac {r_{3+}{\ell} \phi_3 - r_{3-}   t_3}{{\ell}^2 }\,  ,\een
where $r_{3\pm}$ is defined in (\ref{r3pm}).

Note that  a shift in  $T$ is a boost in the Minkowski  $v-y$  plane and a  shift  in $\Phi$  corresponds to a boost in the Minkowski  $u-x$  plane.
Since $\phi_3$  of the BTZ metric (\ref{BTZ}) is periodic with  period $2 \pi$,  the coordinates $\{T, \Phi\}$ must be identified 
under the composition of two discrete boosts:
\ben
\bigl( T, \Phi \bigr ) \,  \rightarrow \, \bigl (T- \frac{2\pi r_{3-}}{\ell}, 
\Phi + \frac{2\pi r_{3+}}{\ell} \bigr ) \, .  \een

\subsection{Kaluza-Klein Reduction of the Scalar Wave Equation}

The  five-dimenskonal Kaluza-Klein metric Ansatz
\be
ds_5^2 = e^{\phi_1} \gamma_{\alpha \beta} dx^{\alpha} dx^{\beta} + e^{-2 \phi_1} ( d{z} + {\cA_2}_{\alpha} dx^{\alpha})^2\, ,
\ee
where $\{\alpha,\beta\}=0,1,2,3$,  results in the five-dimensional wave equation  given by
\be
\nabla^{\alpha} \nabla_{\alpha} {\bf \Phi}  - \nabla^{\alpha}{\cA_2}_{\alpha} \partial_{{z}} {\bf \Phi}  - 2 {\cA_2}^{\alpha} \nabla_{\alpha} \partial_{{z}}
{\bf  \Phi}  + ({\cA_2})^2 \partial_{{z}}^2 {\bf \Phi}  = 
- e^{\phi_1} \partial_{{z}}^2 {\bf \Phi}\, .
\ee
If we  make the assumption that $ {\bf \Phi}  $ is separable in term of a four-dimensional wave function and a function of the fifth coordinate $z$:
\be 
{\bf \Phi} (x^{\alpha}, {z})={\bf \Phi} (x^{\alpha}) e^{ i f(z)}\,, 
\ee
we can rewrite the above equation as
\be
\gamma^{\alpha \beta} \left( \nabla_{\alpha} - 
i (\partial_{{z}} f ){\cA_2}_{\alpha} \right)\left( \nabla_{\beta} - i 
(\partial_{{z}} f ){\cA_2}_{\beta} \right) {\bf \Phi} ( x^{\alpha}) = ( \partial_{{z}} f)^2 e^{\phi_1} {\bf \Phi( x^{\alpha})} \,  .
\ee
For the compactification on a circle $S^1$ with radius $2\pi R$, the above equation is solved with the   Ansatz  for   $f(z)=kz$, where the wave number $k$ is quantised in units of $R^{-1}$. The remaining  effective four-dimensional  wave equation can then  be interpreted  as  the Klein-Gordon equation of the four-dimensional charged particle  with a charge $q=k$ and an effective mass $\propto k$ which is modulated by the scalar field $e^{\phi_1}$:
\be
m_{eff}^2 = k^2 e^{\phi_1} \, .
\ee

\end{document}